\documentclass{iopart}

\usepackage{graphicx}
\begin{document}
\title[Phase-controlled slow light in
current-modulated semiconductors]{Phase-controlled slow and fast
light in current-modulated semiconductor optical amplifiers}
\author{M.A. Ant\'on, F. Carre\~no, \'Oscar G. Calder\'on,
Sonia  Melle, Francisco Arrieta-Y\'a\~nez}
\address{Escuela Universitaria de \'Optica,
Universidad Complutense de Madrid, C/ Arcos de Jal\'on s/n, 28037
Madrid, Spain}
\ead{antonm@fis.ucm.es}
\begin{abstract}
We present a theoretical study of the slow and fast light
propagation in semiconductor optical amplifiers based on coherent
population oscillations. By modulating the injection current to
force the population oscillations we can modify the delay or
advancement of light signals. Specifically, it is shown that the
relative phase of the optical signal to the bias current modulations
can be used as a knob for changing light propagation from delay to
advancement. In addition, we analyze the effect of the modulation
current  for slow light in vertical cavity surface emission lasers
(VCSELs) by taking into account the cavity effects. It is shown that
the change of the depth of the modulation allows to tune the
structural resonance, which in turn produces an enhancement of the
delay.
\end{abstract}


\maketitle


\section{Introduction}\label{sec:INTRO}

Recent dramatic experimental demonstration of slow and fast light
has stimulated considerable interest in the dynamic control of the
group velocity of light and in the development of tunable
all-optical delays for applications such as optical buffers. Two
methods are generally exploited to control optical delay: one of
them relies in the use of dispersive devices and the other is based
on the modification of the group index of an optical medium. The
first approach is structural, where one aims for finding an optimal
structure that enhances the nonlinear response (through its
geometrical properties). Some of the most promising systems that
explore this approach are Fabry-Perot resonators, high Q cavities,
and photonic crystals. The second approach makes use of nonlinear
optical effects such as electromagnetically induced transparency
(EIT) \cite{BOYD02,HAU99,KASH99,WANG00}, coherent population
oscillation (CPO) \cite{Bigelow03b}, Raman \cite{R1,R2} and
Brillouin amplification \cite{B1,B2,B3}. These nonlinear optical
techniques have been extensively studied in atomic systems, however
the corresponding counterparts in solid state crystals, fibers, and
semiconductors are extremely attractive in order to obtain optical
tunable  and easily integrable devices like delay lines, and
buffers. EIT in a solid has been demonstrated with a rare-earth
doped matrix \cite{TR1} and in a Pr-doped solid at $5$ K \cite{TR2}.
Some of the coherent effects which are present in dilute systems
have been also investigated theoretically and experimentally in
semiconductors \cite{SC1,SC2,SC3}. In particular, EIT was
experimentally obtained using exciton and biexciton transitions  in
a quantum-well (QW) structure \cite{A1,A2,A3,A4}.

Coherent population oscillations have been shown to be a robust
physical mechanism which allows for the variation of group velocity.
CPO produces a narrow hole in the absorption or gain profile as a
consequence of the periodic modulation of the ground state
population at the beat frequency between a strong control field and
a weak probe field sharing a common atomic transition. Unlike EIT,
which is dominated by the coherence dephasing time, CPO is governed
by the population relaxation time and becomes nearly insensitive to
temperature. In addition, CPO is weakly influenced by inhomogeneous
broadening in atomic systems in contrast to the quantum coherence
effect involved in EIT. Slow and fast light at room temperature
originated by CPO has been experimentally observed in solid state
crystals \cite{Bigelow03b,Bigelow03,Baldit05}, erbium doped fibers
(EDFs) \cite{Zhang08,Schweinsberg06,Melle07}, photorefractive
materials \cite{ODULOV04,ZHANG04}, and biological thin films
\cite{Wu05}, among others.

Slow and fast light in semiconductor optical amplifiers (SOAs) has
also been studied  extensively in recent years \cite{MORKaa,MORKbb},
because those such systems have the advantage of providing
compactness, easy integration with electronic or optical systems,
large bandwidth due to fast carrier dynamics, and easy and quick
tuning of delay by direct current injection or optical pumping. When
a strong control beam and a weak signal beam (at different
wavelengths) propagate through a SOA, beating between the two beams
causes oscillations of the carrier density. These oscillations
create dynamical gain and index gratings in the device. Interaction
of the signal beam with the dynamical gratings results in the group
index change experienced by the signal. The group index can be
controlled either electrically (by changing the bias current of the
SOA) or optically (by changing the pump power). Using this method, a
group index reduction of $10$ has been demonstrated in a compact $2$
mm device.  CPO based slow light has also been reported in a
multiple-quantum-well structure at low temperature \cite{SC1} and
in quantum-dot (QD) semiconductor optical amplifiers operating at
$1.3\,\mu$m at room temperature \cite{SU1}. Recently, Su {\it et al}
reported that the four-wave mixing (FWM) effect, in conjunction with
the CPO effect, plays an important role in a quantum-well SOA in the
gain regime \cite{SU2}. Furthermore, slow light at room temperature
and with a bandwidth in the range of GHz can be achieved in
semiconductor quantum wells and quantum dots \cite{WEI07}. All these
results can be understood by considering the theory of CPO and FWM
(see \cite{CONIREW07,CONIREW08} and references therein for a
review). It has been shown that the delay or advancement achieved
saturates with both pump and signal powers and is limited by the
carrier lifetime. However, the effects of the refractive index
mediated by wave mixing can be exploited to increase the degree of
light control by optical filtering prior to detection \cite{MORK08}.
In the conventional CPO studies in SOAs, the optical beam is
modulated in the RF range and delay or advancement of the detected
signals are measured depending on the value of the DC bias current,
according to be below or above the transparency current. Therefore,
current modulation is another mechanism which can be externally
controlled to modify the semiconductor optical response.   Here, we
will analyze the possibility of realization of slow and fast light
in a SOA by considering the simultaneous modulation of the optical
beam and the bias current in such a medium. The dynamics of SOAs
subject to direct current modulation has not been dealt in the slow
light context. The feasibility to periodically modulate the bias
current in the RF range has been previously addressed \cite{MORK99}
from a theoretical point of view. The possibility to modify at will
the relative phase of the modulated bias current to the probe field
allows us to manage the magnitude of the delay or advancement
experienced by the probe field.

\section{Theoretical model of forced
population oscillations} \label{sec:MODEL}

\begin{figure}[ht!]
\centering
\includegraphics[scale=0.55]{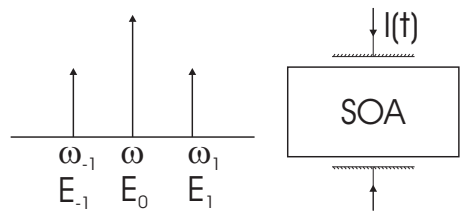}
\caption{Current modulated SOA and spectral components of the
optical field impinging on it. The angular frequencies of the
sidebands are detuned from $\omega$ by $\delta$: $\omega_{\pm
1}=\omega \pm \delta$. \label{fig:setup}}
\end{figure}

We consider a SOA driven by an injected current $I$. A laser field
couples the transition between the semiconductor valence band and
the conduction band. Typical spontaneous carrier lifetimes are in
the order of a few nanoseconds. By adjusting the SOA injection
current we may achieve an amplifying/absorbing semiconductor medium
at relatively low injection currents, $I$, ranging from a few mA to
several hundred of mA.

To model the population oscillation in the semiconductor structure,
the theoretical starting point relies on considering that an optical
beam ${\cal E}(t)$ impinges on the medium. The optical field is
given by

\begin{equation}
{\cal E}(t) = \frac{1}{2} E(t) e^{-i\omega t} + c.c. \; ,
\label{eqcampo1}
\end{equation}

\noindent $\omega$ being the angular frequency of the optical field
and $E(t)$ is the slowly-varying amplitude. We assume that this
field is comprised of a strong DC control beam $E_0$ and two
sidebands $E_1$ and $E_{-1}$ separated by the modulation angular
frequency $\delta$ which lies within the RF range. The explicit
expression for $E(t)$ is given by

\begin{equation}
 E(t) =  E_0(t) + E_{1}(t) e^{-i \delta t} + E_{-1}(t) e^{i \delta t} \; .
\label{eqcampo2}
\end{equation}

\noindent When this modulated beam goes through the SOA (see figure
\ref{fig:setup}), the three components of the electrical field
interact with the carriers in the semiconductor through stimulated
emission and will impose a modulation on the carrier density due to
the frequency beating between the optical waves. When the beating
frequency $\delta$ is small enough that the carriers in the SOA can
follow the oscillation between the valence band and the conduction
band, the carriers will generate a temporal grating and induce the
energy exchange between the control and the sideband fields. This
process creates a spectral hole seen by the sideband fields with
width on the order of GHz (inverse of the carrier lifetime).

\noindent The carrier density $N$ attained for a selected bias
current $I$, is obtained by solving the following rate equation:

\begin{equation}
\label{eqcampo3}
 \frac{d\; N}{d\; t} =
 \frac{I}{q V} - \frac{N}{\tau} -  \frac{1}{2} n_{bg} c \epsilon_0\frac{\Gamma g(N)}{\hbar
\omega_0}  \left<|E(t)|^{2} \right> \; ,
\end{equation}

\noindent where $q$ is the electron charge, $V$ is the active
volume, $g(N)$ is the modal gain experienced by the optical field
when the current $I$ is injected into the active region of the
semiconductor amplifier, and $\Gamma$  is the confinement factor,
that is, the fraction of mode energy confined within the active
volume $V$, $\omega_0$ is the angular frequency between the
semiconductor valence band and the conduction band, $\tau$ is the
carrier lifetime, and $n_{bg}$ is the background refractive index of
the material. The angle brackets denote the averaging operation over
the active volume. Equation (\ref{eqcampo3}) can be derived from the
density-matrix equations \cite{SARGENT,AGRAWAL} in the rate-equation
approximation. In equation (\ref{eqcampo3}) the effects of carrier
diffusion in the transverse direction have been ignored since the
transverse dimensions of the SOA are generally smaller than the
diffusion length. In addition, a linear modal gain $g(N)$ is assumed
to be equal for all the involved waves, an assumption justified
since the pump-probe detuning is much smaller compared to the
gain-spectrum bandwidth, i.e., $\delta \tau \ll 1$, then the
following condition is satisfied:

\begin{eqnarray} \label{eqcampo4}
g(N)=\alpha (N-N_t) \;  ,
\end{eqnarray}

\noindent $\alpha$ being the gain cross-section and $N_t$ is the
carrier density at which the active region becomes transparent.
These equations neglect ultrafast gain nonlinearities like carrier
heating and spectral hole burning, which is a good approximation for
moderate modulation frequencies below 20 GHz \cite{MORK99}.

In the conventional CPO studies in SOAs, the optical beam is
modulated in the RF range and delay or advancement of the detected
signals are measured depending on the value of the DC injection
current, according to be below or above the transparency current.
Here, we will analyze the possibility of improving slow and fast
light performance in SOAs by considering the simultaneous modulation
of the optical beam and the bias current in such a medium. Both
magnitudes are modulated at the same  frequency $\delta$. The
feasibility to produce modulations in the bias current in the RF
range has been previously addressed from a theoretical point of view
\cite{MORK99}.

Our goal is to force the population oscillations by modulating the
bias current $I(t)$, which is described as:

\begin{equation} \label{eqcampo5}
 I(t) =  I_0 + I_{+1} e^{-i(\delta t-\Psi)}+I_{-1}e^{i(\delta t-\Psi)} \; ,
\end{equation}

\noindent where $I_{\pm 1} \ll I_0$, and it is assumed that the
modulated current could be out of phase with respect to the
modulation of the optical weak probe field by a magnitude $\Psi$,
which can be externally changed. In what follows we also assume that
the current is independent on the spatial coordinates, i.e., we will
consider a traveling microwave which will be matched exactly with
the propagating optical fields. When the modulated beam propagates
through the SOA, the three components of the electrical fields
interact with one another and result in a relative phase shift. The
interaction of the sidebands and the control beam causes CPO,
modifies the temporal refractive index, and changes the group
velocity of the light signal. Simultaneously, the injected current
and optical beams are also coupled to each other through the
wave-mixing effects, resulting in an additional phase shift.

To solve equation (\ref{eqcampo3}), we substitute $E(t)$, $g(N)$ and
$I(t)$ from equations (\ref{eqcampo2}), (\ref{eqcampo4})  and
(\ref{eqcampo5}), respectively, which results in the following
equation:

\begin{eqnarray} \label{eqcampo6}
\fl \frac{d\; N}{d\; t} & = & - \frac{N}{\tau} + \frac{N_t}{\tau}
\left( R_0 + R_{1} e^{-i(\delta t-\Psi)} +
R_{-1} e^{i(\delta t-\Psi)} \right) \\
\fl & & - \frac{(N-N_t)}{\tau P_{sat}} \frac{1}{2}n_{bg}c\epsilon_0
\left[ |E_0|^2+ \left( E^{*}_{0} E_{1} + E_{0} E^{*}_{-1} \right)
e^{-i\delta t} + \left(E_{0} E^{*}_{1} + E^{*}_{0} E_{-1} \right)
e^{i\delta t} \right] \; , \nonumber
\end{eqnarray}

\noindent where we have defined the following normalized currents
$R_{0,\pm1}$ and the saturation power $P_{sat}$:

\begin{eqnarray} \label{eqcampo7}
R_{0,\pm 1}& =&  \frac{\tau}{q V N_t} I_{0,\pm 1} \; ,  \\
P_{sat} & = & \frac{\hbar \omega_0}{\Gamma \alpha \tau} \; .
\end{eqnarray}

Next we assume that the carrier density can be described by a DC
term and small AC terms modulated at the same beating frequency,
i.e.,

\begin{equation} \label{eqcampo10}
N(t) =  N_0 + N_{1} e^{-i \delta t}+ N_{-1} e^{i \delta t} \; ,
\end{equation}

\noindent where $N_0$ is the static carrier density and $N_{\pm 1}$
is the amplitude of the carrier population oscillation of the
corresponding sideband. The solution of equation (\ref{eqcampo6})
yields the following expressions for the carrier density amplitudes:

\begin{eqnarray} \label{eqcampo11}
N_0 & = & N_t \; \frac{R_0 + q_0}{1 + q_0} \; ,  \\
N_{1} & = &  N_t \; \frac{R_{1} e^{i \Psi} - (N_{0}/N_t - 1) q_{1}
}{
1 + q_0 - i \delta \tau} \; ,  \\
N_{-1} & = & (N_{1})^{*} \; ,
\end{eqnarray}

\noindent where we have defined the following normalized optical
powers:

\begin{eqnarray} \label{eqcampo7}
q_{0} & = &  \frac{1}{2} n_{bg} c \epsilon_0 \frac{|E_0|^2}{P_{sat}} \; , \\
q_{1} & = &  \frac{1}{2} n_{bg} c \epsilon_0 \frac{\left( E^{*}_{0}
E_{1} + E_{0} E^{*}_{-1} \right) }{P_{sat}} \; .
\end{eqnarray}

\noindent In view of the previous considerations, we arrive at the
following equation for the carrier density oscillation:

\begin{equation} \label{eqcampo12}
N(t)  =  N_0 + N_t \left[ \frac{R_{1} e^{i \Psi} - (N_0/N_t - 1)
q_{1} }{ 1 + q_0 -i \delta \tau} \;  e^{-i \delta t} + c.c. \right]
\; .
\end{equation}

\noindent A close inspection of equation (\ref{eqcampo12}) reveals
that population oscillation works as a temporal grating whose
amplitude depends both on the coupling between the DC and the
sidebands of the probe field, and on the modulation term of the
injection current $R_1$. The oscillating part of equation
(\ref{eqcampo12}) contains two terms: one, which is proportional to
$q_{1}$, is responsible for the conventional population oscillation
observed up to date \cite{AGRAWAL} and arises from the modulation of
the field connecting the optical transition. The other one, which is
proportional to $R_{1}$, arises from the modulation imposed to the
bias current and incorporates explicitly the relative phase $\Psi$.
We will show that this term is responsible for the probe
delay/advancement enhancement.

The response of the system to the weak probe field  can be obtained
by solving the scalar wave equation:

\begin{equation} \label{eqcampo13}
\nabla^{2} {\cal E}(z,t) -\frac{1}{c^2}\frac{\partial^2 {\cal
E}(z,t)}{\partial t^2}= \frac{1}{c^2\epsilon_0}\frac{\partial^2
{\cal P}(z,t)}{\partial t^2} \; .
\end{equation}

\noindent The field given by equation (\ref{eqcampo1}) induces a
complex polarization in the medium

\begin{equation} \label{eqcampo14}
 {\cal P}(z,t) = \frac{1}{2}
 \left( P_0(z) + P_1(z) e^{-i \delta t} + P_{-1}(z) e^{i \delta t}
 \right) e^{-i(\omega t - k z)} + c.c. \; ,
\end{equation}

\noindent where $P_{i}(z)$ ($i=0,1,-1$) is a complex polarization
coefficient that yields index and absorption characteristics for the
DC and side modes waves. It is well-known \cite{AGRAWAL}, that the
induced polarization is given by

\begin{equation} \label{eqcampo15}
{\cal P}(z,t) =   - \frac{c \epsilon_0 (\beta+i) }{\omega
(1-i\Delta)} \alpha \left(N(t)-N_t\right) {\cal E}(z,t) \; ,
\end{equation}

\noindent where $\beta$ stands for the so-called linewidth
enhancement factor and $\Delta=\omega_0-\omega$ is the detuning.
Introducing equation (\ref{eqcampo12}) and (\ref{eqcampo1}) into
equation (\ref{eqcampo15}) allows us to obtain the components of the
polarization $P_{i}(z)$, which are given by

\begin{eqnarray} \label{eqcampo16}
\fl P_{0}(z) & = & - \frac{c \epsilon_0 (\beta+i)\alpha N_t}{\omega
(1-i\Delta) \omega_c} \; \left[ \frac{}{} \left(R_0-1\right) \; E_0 \right.  \nonumber \\
\fl & & \left. +\, \frac{ \left( \omega_c R_{1} e^{i \Psi} - \left(
R_{0} - 1 \right) q_{1} \right) }{ \omega_c - i \delta \tau } \;
E_{-1} + \frac{ \left( \omega_c R_{1} e^{- i \Psi}
- \left( R_{0} - 1 \right) q_{1}^* \right) }{ \omega_c + i \delta \tau } \; E_{1} \right] \; ,  \\
\fl P_{1}(z) & = &  - \frac{c \epsilon_0 (\beta+i) \alpha N_t}{
\omega (1-i\Delta) \omega_c } \left[ \left(R_0 -1\right) \; E_{1} +
\frac{ \left( \omega_c R_{1} e^{i \Psi } - \left(R_{0} - 1
\right) q_{1} \right) }{ \omega_c - i \delta \tau} \; E_0 \right] \; ,  \nonumber \\
\fl P_{-1}(z) & = & - \frac{c \epsilon_0 (\beta+i) \alpha N_t }{
\omega (1-i\Delta) \omega_c}  \left[\left( R_0 - 1 \right) \; E_{-1}
+ \frac{ \left( \omega_c R_{-1} e^{ - i \Psi } - \left( R_{0} - 1
\right) q^{*}_{1} \right) }{ \omega_c + i \delta \tau } \; E_0
\right] \; . \nonumber
\end{eqnarray}

\noindent We have defined the dimensionless frequency
$\omega_c=1+q_0$ which roughly measures the linewidth of the
transparency hole created in the absortion/gain spectrum due to CPO.
Now we substitute equations (\ref{eqcampo16}) in equation
(\ref{eqcampo13}), and by equating the coefficients oscillating at
the same frequency, we arrive at the following set of equations for
the amplitudes of the optical fields in the SVEA approximation:

\begin{eqnarray}\label{eqcampo17}
\fl \frac{\partial E_0}{\partial z} & = & \frac{\alpha N_t(1
-i\beta)}{2 \; \omega_c} \left[\left(R_0-1 \right) \; E_0 + \frac{
\left( \omega_c R_{1} e^{i \Psi} - \left(R_{0} -1\right)
q_{1} \right) }{\omega_c -i\delta \tau} \; E_{-1} \right. \nonumber \\
\fl & & \left. +\, \frac{ \left( \omega_c R_{1} e^{-i\Psi} - \left(
R_{0}- 1 \right) q_{1}^* \right)}{
\omega_c + i \delta \tau}  \; E_{1} \right] \; ,  \\
\fl \frac{\partial E_1}{\partial z} & = & \frac{\alpha N_t(1
-i\beta)}{2 \; \omega_c} \left[ \left( R_0 - 1 \right) \; E_1 -
\frac{\left( R_{0} - 1 \right) q_{1} }{\omega_c -i\delta \tau} \;
E_0 + \frac{ \omega_c R_{1} e^{i \Psi}}{ \omega_c -i\delta \tau}
\; E_0 \right] \; , \nonumber \\
\fl \frac{\partial E_{-1}}{\partial z} & = & \frac{\alpha N_t(1
-i\beta)}{2 \; \omega_c} \left[ \left(R_0 - 1 \right) \; E_{-1} -
\frac{ \left(R_{0}- 1 \right) q^{*}_{1}}{\omega_c+i\delta \tau} \;
E_0 + \frac{\omega_c R_{1} e^{-i\Psi}}{\omega_c +i\delta \tau} \;
E_0 \right] \nonumber \; ,
\end{eqnarray}

\noindent where we have assumed  the probe field to be at resonance
for simplicity, i.e., $\Delta=0$. It is well known that the first
two terms appearing in the equation of evolution of the sidebands
lead to coherent dips in pump-probe spectroscopy \cite{AGRAWAL}. The
first one is related to the linear susceptibility of each sideband,
while the second term arises from the multiple wave mixing process.
This second term is responsible for the creation of a hole in the
probe gain and therefore is the physical origin of conventional CPO.
Here, we should notice that the last term that contributes to the
development of the sidebands arises from a net exchange from the DC
component of the optical field which in turn arises as a consequence
of the modulation of the current ($R_1$). In the case that $\Psi=0$,
the contribution of this term will produce an in phase contribution
to the index and gain gratings that will result in an
enhancement/fall of the CPO depending on the value of the bias
current to be below or above the transparency level. In the case
that $\Psi=\pi$, it will produce a change that will turn delay into
advancement. In the general case that $\Psi \neq 0,\pi$, the last
term will result in a mixing of the gain grating and the index
grating which is responsible for the enhancement of the phase
delay/advancement. Note that the equation for the DC component also
incorporates the effects of the weak sidebands as source terms.

These effects can become more transparent by considering the spatial
propagation of the magnitude $q_1$ which accounts for detection of
the output modulated signal. By neglecting the spatial variation in
$E_0$ (non-depleted approximation) and $\beta=0$ we arrive to the
following expressions for its modulus and phase:

\begin{eqnarray} \label{eq:delay}
\fl \frac{d  \left|q_1\right|}{d \; z} & = & - \frac{\alpha N_t
q_0}{ 2 \; \omega_c} \left[ \left(R_0-1\right) \frac{
\left|q_1\right| }{q_0} \left( 1 - \frac{ 2 \; q_0 \; \omega_c }{
\omega_c^2 + (\delta \tau)^2  } \right) \right. \nonumber \\
\fl & & \left. + \frac{2 \; \omega_c \; R_1}{ \omega_c^2 + (\delta
\tau)^2 } \left( \omega_c \cos\Psi -
\delta \tau \sin\Psi \right) \right] \; , \\
\fl \frac{d \; \phi}{d \; z} & = & - \frac{\alpha N_t q_0}{ \omega_c
\left[ \omega_c^2 + (\delta \tau)^2 \right]} \left[  \left( R_0 - 1
\right) \delta \tau - \frac{\omega_c R_1}{\left|q_1\right|} \left(
\omega_c \sin\Psi + \delta \tau \cos\Psi \right) \right] \; ,
\nonumber
\end{eqnarray}

\noindent  A close inspection of equation (\ref{eq:delay}) reveals
that the four-wave mixing effect cancels out in this detection
scheme. In this case, and in the absence of current modulation
($R_1=0$), the slow-down effect reduces to a saturation phenomenon.
However, the non null value of current modulation will produce a
resonance behavior which mimics the effects of a four-wave mixing
process.

It is worth mentioning that equation (\ref{eq:delay}) reduces to
those obtained by M{\o}rk {\it et al} \cite{MORK99}, with the proper
identifications of variables, except for the fact that in our case
we consider input optical fields at the entrance of the medium (at
$z=0$).

\section{Effect of current modulation on
the phase delay/advancement}

\begin{figure}[ht!]
\centering
\includegraphics[scale=0.4]{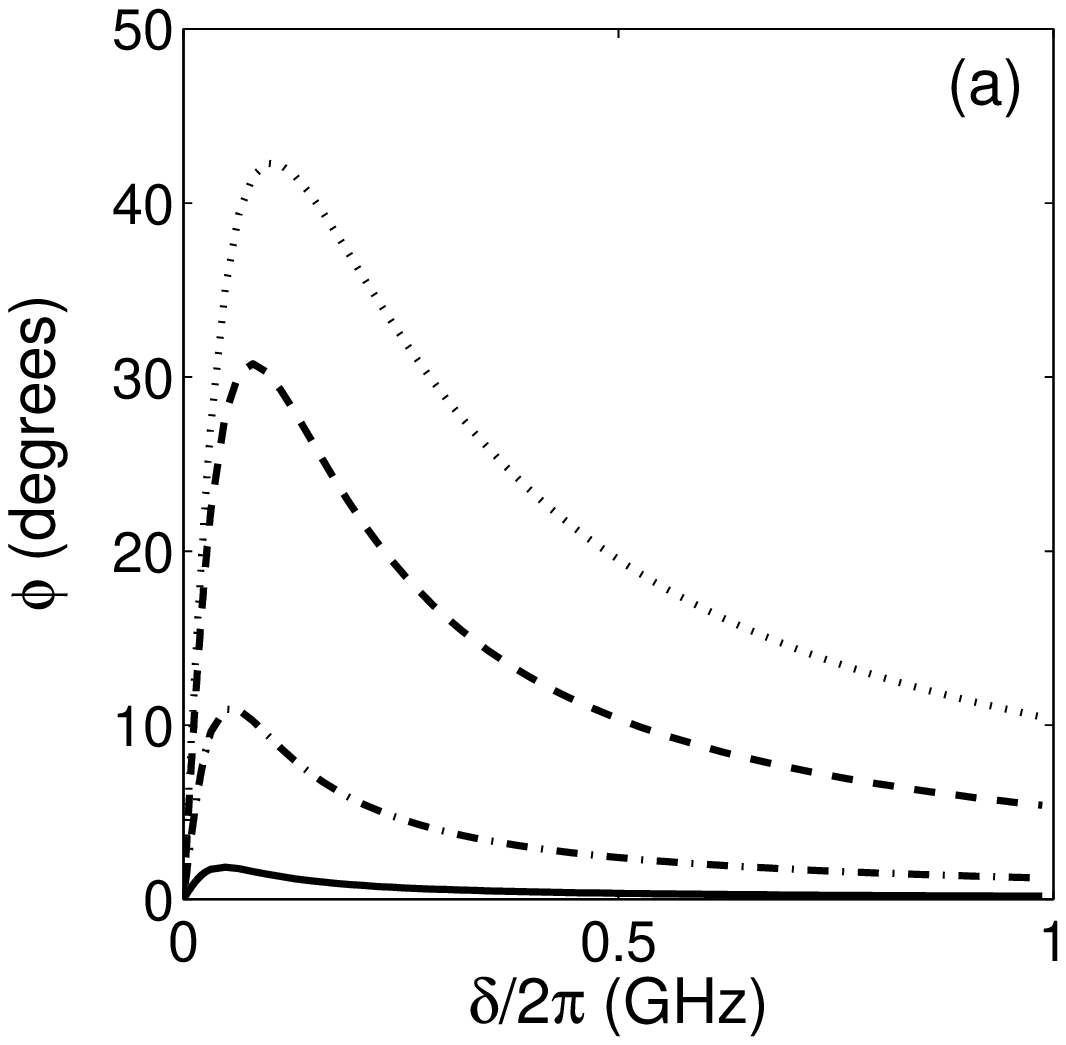}
\includegraphics[scale=0.4]{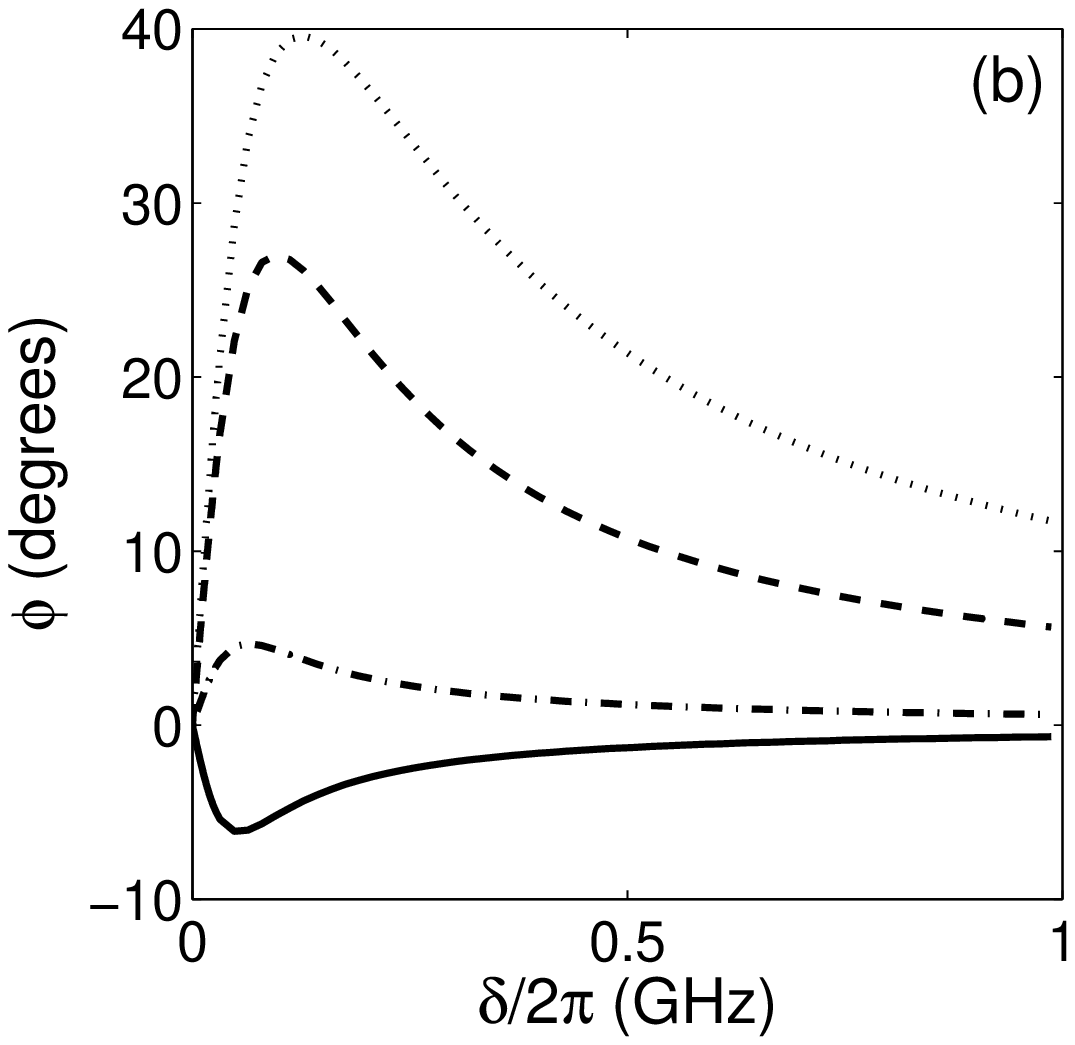}
\caption{Phase delay versus the modulation frequency for different
values of the modulation depth: $R_{1}=0$ (solid line),
$R_{1}=0.01\times R_0$ (dashed-dotted line), $R_{1}=0.05\times R_0$
(dashed line), and $R_{1}=0.1\times R_0$ (dotted line). DC injection
current (a) $R_{0}=0.95$ (b) $R_{0}=1.15$. Other parameters are:
$q_0=0.5$, $q_1=0.1\times q_0$, and $\Psi=0$. \label{figure2} }
\end{figure}

In this section we will perform numerical simulations concerning the
influence of current modulation on slow and fast light in a SOA. For
this purpose we will carry out numerical integration of equation
(\ref{eqcampo17}) for selected values of the parameters involved,
which may be accessible from an experimental point of view. To
illustrate the effect of the current modulation we use the following
parameters: the overlap factor $\Gamma$ is assumed to be $1$, the
linewidth enhancement factor $\beta=0$, the effective carrier
lifetime $\tau=5$ ns, the transparent carrier density $N_t = 1
\times 10^{18}$ cm$^{-3}$, and the length $L= 0.3$ mm. We also
assume that the gain coefficient is $\alpha = 2 \times 10^{-16}$
cm$^{2}$, and use $n_{bg}=3.2$ as the background refractive index.
By considering a typical area in the order of $10^{-12}$ m$^2$, the
saturation power is $P_{sat} = 4.27$ mW. The theoretical operation
of the SOA should have a threshold current (transparency current) in
the order of $I_{t}= 6.4$ mA.

We present in figure \ref{figure2}(a) the results obtained for the
case in which the injection current is below the transparency
current ($R_0<1$). In this case the overall effect of the modulation
current is to produce a huge increase of the maximum phase delay for
moderate values of the modulation while producing a slight increase
of the bandwidth in the range of $1$ GHz. The most remarkable effect
is obtained when operating above the threshold injection current
($R_0>1$) as it is shown in figure \ref{figure2}(b). There, we
appreciate that for a moderate value of the modulation current
(dashed-dotted curve) delay is achieved. The delay tends to increase
over all the range of frequencies with regard to that obtained for
the DC injection current case (solid line). A further increase of
the modulation current turns gain into absorption which reflects in
the obtention of phase delay over all the range of frequencies
(dashed and dotted lines). Thus, the level of modulation of the
current allows the control of the level of advancement and to switch
from fast to slow light. In order to qualitatively explain this
behavior we can use the trivial solution of equation
(\ref{eq:delay}) for the phase delay by neglecting the
$z$-dependence of the rest of magnitudes:

\begin{equation} \label{eq:delay2}
\fl \phi(z=L) \simeq  - \frac{\alpha L N_t q_0}{ \omega_c \left[
\omega_c^2 + (\delta \tau)^2 \right]} \left[  \left( R_0 - 1 \right)
\delta \tau - \frac{\omega_c R_1}{\left|q_1\right|} \left( \omega_c
\sin\Psi + \delta \tau \cos\Psi \right) \right] \; .
\end{equation}

\noindent In the case that $\Psi=0$ (in phase case), this phase
delay (\ref{eq:delay2}) is proportional to $R_0-1- \omega_c R_1 /
|q_1|$, therefore, the current modulation contributes to produce
delay. Thus, for an injection current below the transparency one,
where delay is expected, the current modulation increases its
magnitude. However, at an injection current above the transparency
one, where advancement is expected, the current modulation produces
a decrease of the advancement, and for a threshold value of the
modulation amplitude $R_1 = (R_0-1) |q_1| / \omega_c$, the phase
advancement turns into delay. Therefore, the modulation amplitude
$R_1$ can be used as a control parameter to switch the propagation
regime.

\begin{figure}[ht!]
\centering
\includegraphics[scale=0.55]{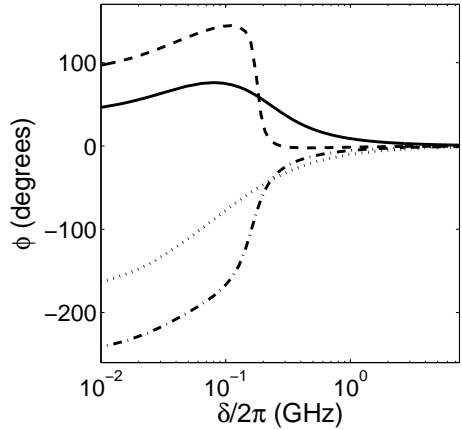}
\caption{Phase delay  versus the modulation frequency for different
values of the relative phase $\Psi$: $\Psi=45^0$ (solid line),
$\Psi=100^0$ (dashed line), $\Psi=120^0$ (dashed-dotted line),
$\Psi=180^0$ (dotted line). Operating point $R_0=0.95$, subject to a
modulation $R_1=0.1\times R_0$. Other parameters as in figure
\ref{figure2}. \label{fig:fig3}}
\end{figure}

In order to show how the relative phase shift of the modulation
current to the modulated optical field ($\Psi$) influences the phase
shift $\phi$, we plot in figure \ref{fig:fig3} the phase delay
versus the beat frequency for several values of $\Psi$. We have
selected an operating point below the transparency current
($R_0<1$). Note that under these circumstances, by solely modifying
the AC current ($R_1$) and by keeping $\Psi=0$, we remain in the
absorptive regime which always produces delay on the optical signal
[see figure \ref{figure2}(a)]. However, when the modulation current
is fixed while the relative phase is properly changed, a switch from
absorption to gain is produced which results in turning delay into
advancement of the optical signal. In other words, the phase $\Psi$
may be used as an external parameter to control the magnitude of
phase delay/advancement achieved for all the range of frequencies
while keeping constant the rest of parameters. The change in the
regime of propagation has its counterpart in the change from
absorption to gain (not shown). This behavior can be easily
explained in terms of the new contributions to the index and gain
gratings originated by the modulation current [see equation
(\ref{eq:delay})]. For the values $\Psi=0$, and $\Psi=\pi$, the
effect of current modulation only affects the absorptive grating,
while for values like $\Psi=\pi/4$ (dashed-dotted line), both the
gain and the index gratings should contribute to the phase delay
experienced by the optical signal. From equation (\ref{eq:delay2}),
we can estimate a threshold value of $\Psi$ to switch from delay to
advancement, i.e., $\Psi \simeq - \arctan \left( \delta \tau /
\omega_c \right) + \arcsin \left[ (R_0-1) |q_1| \delta \tau / (
\omega_c R_1) \right]$.

\begin{figure}[ht!]
\centering
\includegraphics[scale=0.5]{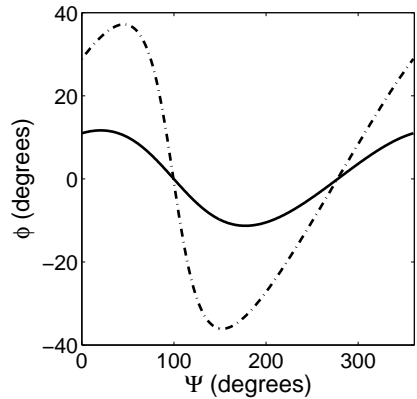}
\caption{Phase delay versus the relative phase $\Psi$ for two values
of the length of the active region: $L=0.1$ mm (solid line) and
$L=0.3$ mm (dashed-dotted line), subject to a modulation
$R_{1}=0.1\times R_0$ for a  modulation frequency $\delta / (2\pi) =
0.3$ GHz. Other parameters as in figure \ref{figure2}.
\label{fig:fig4}}
\end{figure}

Figure \ref{fig:fig4} shows examples of the calculated phase delay
versus the relative phase $\Psi$ for two different values of the
amplifier length ($L$). In the case of small length (solid line),
the phase delay exhibits a nearly sinusoidal behavior. This fact is
consistent with the linear approximation which leads to equation
(\ref{eq:delay2}). An increase of the length of the medium results
in the breaking of the sinusoidal shape (see dashed-dotted line).
This effect arises as a consequence of propagation effects
associated to the numerical solution of equation (\ref{eqcampo17}).

In the previous calculations we have used a null value for the
linewidth enhancement factor ($\beta$). The influence of such
parameter in the index change and the probe gain has been a subject
of analysis in the pioneering work of Agrawal \cite{AGRAWAL}. In
that work, it was shown that the increment of $\beta$ will result in
a distortion of the line shape. In the present case, where we are
interested in the phase of the modulated signal intensity, the
influence of $\beta$ on $\phi$ is unnoticeable, since the coherent
population effect due to the pump beam dominates, while the four
wave mixing effects cancel out.

\subsection{Slow and fast light in the sidebands by optical filtering}

\begin{figure}[ht!]
\centering
\includegraphics[scale=0.4]{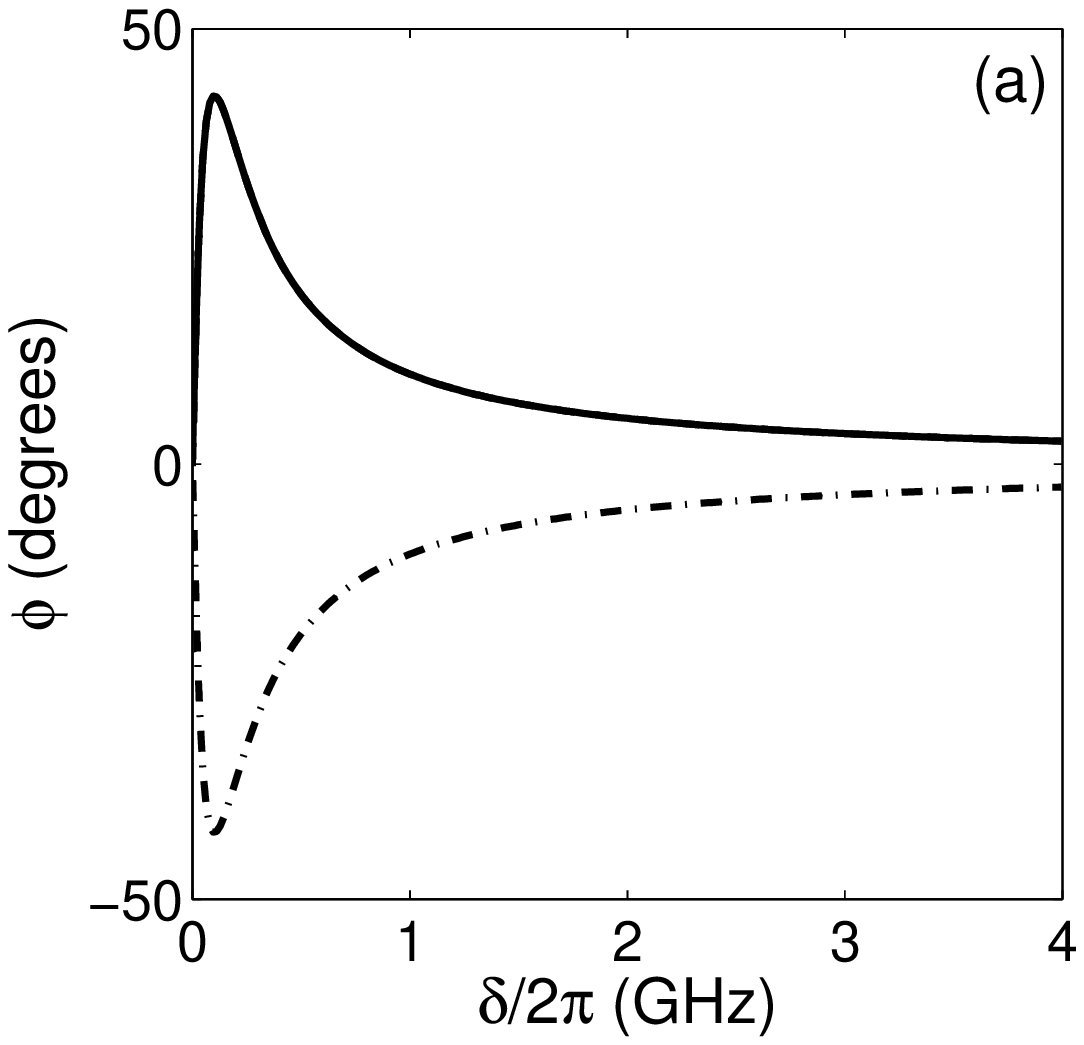}
\includegraphics[scale=0.4]{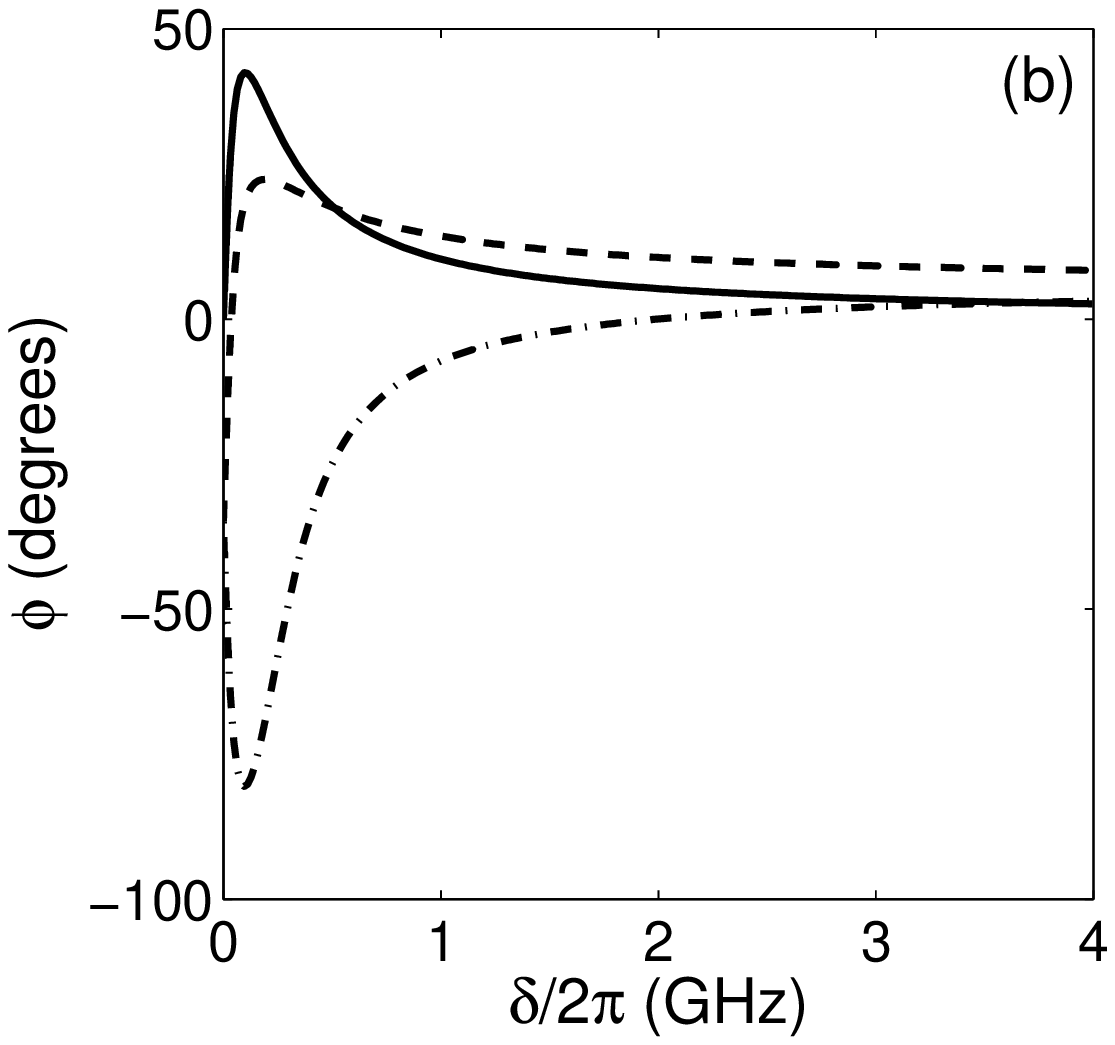}
\caption{Phase delay versus the modulation frequency with $\beta=0$
(a) and with $\beta=1$ (b). Solid line (without optical filtering
$q_1$), dashed-dotted line (by filtering $E_{-1}$), and dashed line
(by filtering $E_{1}$), at a fixed value of the modulation depth
$R_1=0.1$ and DC injection current $R_0=0.95$, and $\Psi=0$. Other
parameters as in figure \ref{figure2}. \label{fig:fig5}}
\end{figure}

In the conventional detection schema commonly used in  many slow
light experiments, no optical filtering is performed in the output
beam of the SOA. However, Xue {\it et al} \cite{MORK08} have shown
that the degree of control of the optical delay in a SOA can be
improved by the application of an optical filtering of the output
signal prior to detection. In the present case, we will also analyze
how the current modulation will affect the delay or advancement when
optical filtering is performed. Figure \ref{fig:fig5}(a) presents
the results obtained for a null linewidth enhancement factor
($\beta=0$). There we appreciate that the blue-filtered signal
(dashed line, filtering $E_1$) nearly coincides with the signal
obtained without optical filtering (solid line), while the
red-filtered signal (dashed-dotted line,  filtering $E_{-1}$)
exhibit advancement in contrast to the other two cases. The three
curves exhibit a symmetric line shape. This behavior can be
explained by simply considering that the four wave mixing terms does
not cancel out. The situation is dramatically modified when
considering a non-null value of $\beta$ as it is displayed in figure
\ref{fig:fig5}(b). In this case the blue-filtered, the red-filtered
and the non filtered values for $\phi$ obtained differ among them,
due to the combined effect of the linewidth enhancement factor and
the modulation current term which is proportional to $R_1$ [see
equation (\ref{eqcampo17})]. Numerical simulations carried out (not
shown) reveal that the greater the value of $\beta$, the greater the
magnitude of the phase delay/advancement obtained. These results
agree with the ones obtained in \cite{AGRAWAL} where the magnitude
of the asymmetry in the index and gain change for the probe was
shown to be $\beta$-dependent.

\begin{figure}[ht!]
\centering
\includegraphics[scale=0.55]{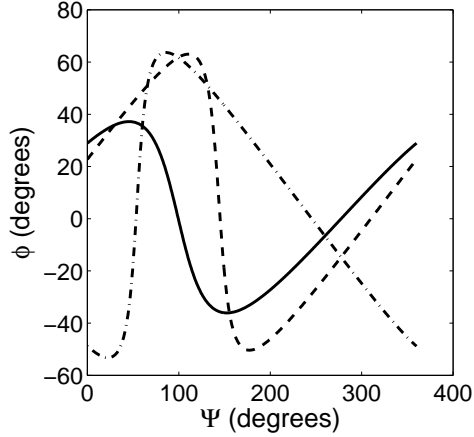}
\caption{Phase shift $\phi$ versus relative phase $\Psi$. Solid line
(without optical filtering $q_1$), dashed-dotted line (by filtering
$E_{-1}$), and dashed line (by filtering  $E_{1}$), at a fixed value
of the modulation depth $R_1=0.1\times R_0$, $\delta/2\pi=0.3$ GHz,
DC injection current $R_0=0.95$, and $\beta=1$.  \label{fig:fig6}}
\end{figure}

Now we fix the detuning to $\delta/2\pi=0.5$ GHz and will allow for
a change of the relative phase $\Psi$. The results are displayed in
figure \ref{fig:fig6} for the three possible cases of the output
signals. It is worth noting that the peak values of
delay/advancement achieved for the filtered signals are greater than
the ones obtained for the non-filtered case. This result resembles
that obtained in \cite{MORK08} in the sense that in the mentioned
paper the authors showed that the filtered signals were shown to
exhibit greater delays than the non-filtered signal, although in
that case the authors changed the input power while we keep the
input power fixed in our simulations.

\begin{figure}[ht!]
\centering
\includegraphics[scale=0.55]{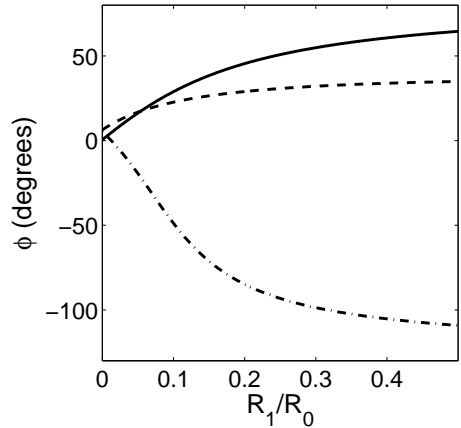}
\caption{Phase shift $\phi_s$ versus $R_1/R_0$. Solid line (without
optical filtering $q_1$), dashed-dotted line (by filtering
$E_{-1}$), and dashed line (by filtering  $E_{1}$) at  a fixed of
$\Psi=0$. Other parameters as in figure \ref{fig:fig6}.
\label{fig:fig7}}
\end{figure}

Figure \ref{fig:fig7} presents the results obtained  versus the
depth of the modulation ($R_{1}$) at the same fixed detuning as
before but now the phase of the modulating current to the modulated
field is also fixed at $\Psi=\pi$. We appreciate that the unfiltered
signal shows a switch from advancement to delay whereas the filtered
signals for both sidebands exhibit the largest delay for the highest
values of the modulating current, leading to a saturation effect for
values larger than $0.4$. For other values of the relative phase
$\Psi$ the behavior is similar to that displayed in figure
(\ref{fig:fig7}) although the level of phase delay/advancement
remains within the same limiting values.

\subsection{Cavity enhanced slow light in a VCSEL}

\begin{figure}[ht!]
\centering
\includegraphics[scale=0.55]{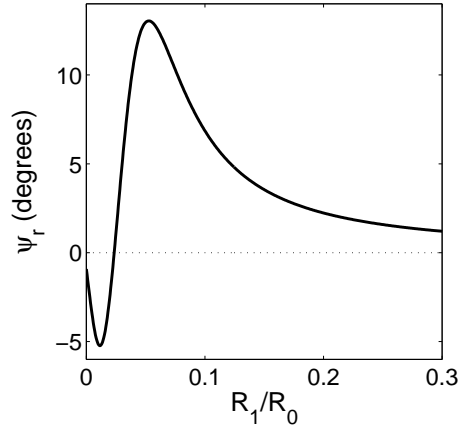}
\caption{Phase delay $\psi_r$ at the output of the VCSEL versus
modulation depth $R_1/R_0$ at a fixed modulation frequency of
$\delta/2\pi=0.05$ GHz, DC injection current $R_{0}=0.95$, $\Psi=0$,
and $\beta=0$.  \label{fig:fig8}}
\end{figure}

Tunable delay using VCSELs has been demonstrated \cite{Zhao05}. In
addition, the temporal dynamics of VCSELs subject to direct current
modulations have also received attention. Verschaffelt {\it et al}
\cite{Verschaffelt} have addressed how the current modulation forces
the population oscillations. Here we turn our attention to extend
our previous model to the case of such a device. The  high
reflectivities of the mirrors in the VCSEL will require to include
the cavity effects in the optical response of the medium together
with the population oscillations.  A simple way to incorporate both
effects relies in the consideration of the VCSEL as a Fabry-Perot
filter with a gain per pass. This approach can be justified by
considering the small longitudinal dimensions of the device which
are in the order of a few micrometers. For the sake of simplicity,
we will also consider that the effects of the intrinsic
birefringence of the medium  can been neglected, thus we can use the
scalar wave equations given in  equation (\ref{eqcampo17}). This
approach has been adopted in several studies concerning slow light
in these devices \cite{LAURAND,CHIA-SHENG}. In view of the previous
considerations, the reflectance $G_r$ obtained for the filter, is
given by \cite{LAURAND}
\begin{eqnarray}
G_r & = & \frac{(\sqrt{R_t}+\sqrt{R_b}\,g_s)^2 +
4\sqrt{R_t}\sqrt{R_b}\,g_s\sin^2 \left[phi\right] }
{\left(1-\sqrt{R_tR_b}\,g_s \right)^2 +
4\sqrt{R_t}\sqrt{R_b}\,g_s\sin^2 \left[\phi\right]}\;,
\end{eqnarray}
where $R_b/R_t$ stands for the bottom/top mirror reflectance, $g_s$
is the single-pass gain, and $\phi$ is the single-pass phase delay
obtained. The magnitudes $g_s$ and $\phi$ are obtained from the
numerical solution obtained from equation (\ref{eqcampo17}). Figure
\ref{fig:fig8} presents the numerical results obtained for the phase
of $G_r$, $\psi_r$,  by considering an active region of length
$L=1.13\,\mu$m. The reflectances of the top and bottom mirrors are
assumed to be $R_t=0.997$, and $R_b=0.99$, respectively.  Note that
for this particular modulation frequency, the changes in the depth
of the bias current modulation results in changes from one regime of
propagation to the other. The most remarkable feature seen in figure
\ref{fig:fig8} is the appearance of a  resonance whose origin is
attributed to the tuning of a mode in the Fabry-Perot filter. The
change of the depth of the modulation allows to tune the resonance,
which in turn produces an enhancement of the delay.

\section{Conclusions}

 In this work we have presented numerical simulations concerning to
 the enhancement of the delay/advancement based in CPO in SOA when
 the bias current is modulated at the same beating frequency.
 An overall increase of delay/advancement is obtained for all
 the frequency range. The depth of the modulation, and the
 reference phase  are shown to have a dramatic influence on
 the magnitude of the phase delay. The slow and fast light have been
 also considered for the case of filtering  the output optical
 fields prior to the detection.  Significant changes in the phase delay are
 predicted by adjusting the relative phase or the modulation depth
 for the filtered signals. The model is also extended to a VCSEL by adding
 the cavity effects arising from the hight reflectivities Bragg mirrors
 of the device.  In this case, it is also shown that the modulation depth
 can be used to tune the structural resonance of the device modeled as a
 Fabry-Perot filter, which results in a phase delay of the  output signal.

\ack

This work has been supported by Projects no. PR34/07-15847
(UCM/BSCH), FIS2007-65382 (MEC), CCG07-UCM/ESP-2179 (UCM-CM),
CCG08-UCM/ESP-4220 from Spain.

\end{document}